\documentclass[11pt]{article}

\usepackage{amsmath,amssymb}

\usepackage{epsfig}
\usepackage{graphicx}               
\usepackage{url}
\usepackage{hyperref}
\usepackage{slashed}
\usepackage{color}
\usepackage{cite}

\newcommand{\nc}{\newcommand}

\nc{\beq}{\begin{equation}}
\nc{\eeq}{\end{equation}}
\nc{\beqa}{\begin{eqnarray}}
\nc{\eeqa}{\end{eqnarray}}
\nc{\bea}{\begin{eqnarray}}
\nc{\eea}{\end{eqnarray}}
\nc{\ra}{\rightarrow}
\nc{\lsim}{\begin{array}{c}\,\sim\vspace{-21pt}\\< \end{array}}
\nc{\gsim}{\begin{array}{c}\sim\vspace{-21pt}\\> \end{array}}
\nc{\Tr}{{\rm Tr}}
\nc{\slsh}{\slash\hspace*{-0.22cm}}

\def\be{\begin{equation}}
\def\ee{\end{equation}}
\def\bea{\begin{eqnarray}}
\def\eea{\end{eqnarray}}
\def\bit{\begin{itemize}}
\def\eit{\end{itemize}}

\title{
\vspace*{-2.3cm}
\begin{flushright}
\normalsize{
  }
\end{flushright}
\vspace{1.5cm}
\Large
\textbf{
Chromo-Rayleigh Interactions of Dark Matter
}
\vspace*{1.0cm}
}

\author{{\bf Yang Bai} and {\bf James Osborne}
\vspace{5mm}
\\
 \normalsize\emph{Department of Physics, University of Wisconsin-Madison, Madison, WI 53706, USA}  
}

\date{}

\begin{document}
\setcounter{page}{0}
\maketitle

\vspace*{1cm}
\begin{abstract}
 \normalsize{
 For a wide range of models, dark matter can interact with QCD gluons via chromo-Rayleigh interactions. We point out that the Large Hadron Collider (LHC), as a gluon machine, provides a superb probe of such interactions. In this paper, we introduce simplified models to UV-complete two effective dark matter chromo-Rayleigh interactions and identify the corresponding collider signatures, including four jets or a pair of di-jet resonances plus missing transverse energy. After performing collider studies for both the 8 TeV and 14 TeV LHC, we find that the LHC can be more sensitive to dark matter chromo-Rayleigh interactions than direct detection experiments and thus provides the best opportunity for future discovery of this class of models.
}
\end{abstract}

\thispagestyle{empty}
\newpage

\setcounter{page}{1}

\baselineskip18pt

\vspace{-2cm}
\section{Introduction}
\label{sec:intro}
Although there is no doubt that dark matter interacts gravitationally both among itself and with the Standard Model (SM) particles, we still have no convincing evidence for other dark matter interactions. Among the three forces in the SM, dark matter can not have order-one couplings under the electroweak forces. Otherwise, it will either emit photons or scatter off nuclei with a too-large cross section in direct detection experiments. Suppressed couplings to the $Z$ boson or the Higgs boson will be probed further in upcoming dark matter experiments~\cite{Cahill-Rowley:2014boa}. For QCD interactions, the dark matter particle can not have a color charge due to confinement. If, instead, it is a QCD composite particle, the hadronic bound states should have a large interaction strength with SM pions and thus have a large scattering cross section with nucleons. This type of dark matter can not penetrate the Earth to reach detectors in underground direct detection experiments, but satellite-based X-ray quantum calorimetry experiments impose stringent bounds on this scenario~\cite{Erickcek:2007jv}.

Another way for dark matter to interact with gluons is through effective contact interactions, comparable to photon--molecule Rayleigh interactions. Introducing such higher-dimensional effective operators to describe dark matter \emph{chromo-Rayleigh} interactions is then independent of whether the dark matter is elementary or composite. For instance, in supersymmetric models the neutralino can couple to two gluons via a stop--top quark loop. In extra-dimensional models, a similar loop from the top quark and its Kaluza-Klein mode can generate the effective chromo-Rayleigh dark matter interaction~\cite{Dobrescu:2007ec}. Composite dark matter models can also generate chromo-Rayleigh interactions, with strengths typically suppressed by the composite scale~\cite{Bagnasco:1993st}. In this paper, we perform a generic study on dark matter chromo-Rayleigh interactions. We pay particular attention to collider searches and the potential signatures associated with chromo-Rayleigh interactions. This study is similar to searches for electrical Rayleigh interactions of dark matter in Ref.~\cite{Frandsen:2012db,Liu:2013gba}, where the electrically charged particles can be searched for directly at the LHC.

For simplicity, we choose the dark matter particle to be a complex scaler denoted as $X$. We study two types of dimension-six interactions: $X^\dagger X \, G^2$ and $\left (X X - X^\dagger X^\dagger \right ) G \widetilde{G}$, where $G$ is the QCD field tensor. For direct detection searches, the first operator provides spin-independent scattering while the second operator provides spin-dependent and momentum-suppressed scattering. For collider searches, the standard model independent signature is a mono-jet plus missing transverse energy. Taking into account the well-known limitations of such searches~\cite{Chang:2013oia,An:2013xka,Bai:2013iqa,DiFranzo:2013vra,Buchmueller:2013dya,Batell:2013zwa,Papucci:2014iwa,Bai:2014osa,Chang:2014tea,Agrawal:2014ufa,Garny:2014waa,Yu:2014pra,Buckley:2014fba,Berlin:2015wwa,Jacques:2015zha,Garny:2015wea,Chala:2015ama,Bell:2015sza}, we introduce simplified models to UV-complete these two operators. Collider signatures owing to such UV-completions depend strongly on how the colored mediators decay. We discuss several possible signatures, including multi-jets and pair-produced di-jet resonances with missing transverse energy.

The rest of this paper is organized as follows. In Section~\ref{sec:contact}, we work out the dark matter thermal relic abundance, direct detection cross sections, and collider constraints from the effective chromo-Rayleigh contact interactions. In Section~\ref{sec:simplified-model}, we introduce two simplified models to UV-complete the previous contact operators. Section~\ref{sec:colored-O1} introduces a scalar color-octet with the signature of four jets plus missing transverse energy, Sections~\ref{sec:colored-O2}--\ref{sec:collider-pair-dijet-met} introduce two fermion color-triplets with the additional signature of paired di-jet resonances plus missing transverse energy, and Section~\ref{sec:compare} offers a comparison to direct detection experiments. We conclude in Section~\ref{sec:conclusion}.

\section{Contact Interactions}
\label{sec:contact}
Our model consists of a complex scalar dark matter field, $X$, which is a singlet under the SM gauge group. We introduce the following two $CP$-conserving, dimension six operators coupling dark matter to the gluon field
\beqa
{\cal O}^{\rm cRayleigh}_1 &=& \frac{\alpha_s}{4\, \pi\,\Lambda^2_1} X^\dagger X G^a_{\mu \nu} G^{a\, \mu \nu}  \,, 
\label{eq:operator1}\\
{\cal O}^{\rm cRayleigh}_2 &=& \frac{i\,\alpha_s}{4\, \pi\,\Lambda^2_2} (X X - X^\dagger X^\dagger) G^a_{\mu \nu} \widetilde{G}^{a\, \mu \nu} \,,
\label{eq:operator2}
\eeqa
where $\Lambda_i$ is the cutoff scale and $\widetilde{G}^{a\, \mu \nu} = \frac{1}{2}\, \epsilon^{\mu\nu\alpha\beta} G^a_{\alpha \beta}$ is the dual gluon field strength tensor. The overall operator normalization accounts for a loop factor. In order for $X$ to be stable, we impose a $\mathcal{Z}_2$ symmetry under which $X$ is odd. Based on these two effective operators, we first calculate the thermal relic abundance, direct detection cross sections, and collider constraints.

\subsection{Thermal Relic Abundance}
\label{sec:contact-thermal}
Depending on the UV physics, the dark matter sector could be more complicated than just one state. Therefore, the dark matter thermal relic abundance calculation based entirely on the operators in Eqs.~(\ref{eq:operator1}) and (\ref{eq:operator2}) can only provide a guidance for the potential parameter space in $M_X$ and $\Lambda_i$ for thermal dark matter. For the first operator, we have the dark matter self-annihilation rate from the process $X^\dagger X \rightarrow G\, G$ as
\beqa
\frac{1}{2}\left[ \langle \sigma v \rangle ( X^\dagger X \rightarrow G\, G) \right] = \frac{1}{2} \left[ \frac{\alpha_s^2}{\pi^3}\, \frac{M_X^2}{\Lambda_1^4} \right]  \equiv s \,,
\eeqa
to leading order in the dark matter's relative velocity $v$ expansion. Here, the overall factor of $1/2$ is due to the relic density being comprised of particles and antiparticles. For the second operator, we have the annihilation rate from the process $XX[X^\dagger X^\dagger] \rightarrow G\, G$, in terms of components with $X = (X_R + i X_I)/\sqrt{2}$, as
\beqa
\frac{1}{2}\left[ \langle \sigma v \rangle ( X_R X_I \rightarrow G\, G) \right] =  \frac{1}{2} \left[ \frac{\alpha_s^2}{\pi^3}\, \frac{M_X^2}{\Lambda_2^4} \right] \equiv s\,.
\eeqa
The dark matter relic abundance is inversely proportional to the annihilation rate and has a formula
$\Omega_X h^2 \approx 1.07 \times 10^9\,\mbox{GeV}^{-1}\,x_F/(\sqrt{g^*}\,M_{\rm pl}\, s)$ with $g^*$ as the number of relativistic degrees of freedom at the freeze-out temperature and is taken to be 86.25 and the Planck scale is $M_{\rm pl} = 1.22 \times 10^{19}$~GeV. The freeze-out temperature $x_F$ is given by $x_F = \ln{\left[ 0.05\,g\,M_{\rm pl} M_X s /(\sqrt{g^*\,x_F} ) \right]}$ with $g=2$ and  is typically ${\cal O}(20)$.

\subsection{Direct Detection}
\label{sec:contact-direct}
For the first operator ${\cal O}^{\rm cRayleigh}_1$, we can use the matrix element of $G^a_{\mu \nu} G^{a\, \mu \nu}$ inside a nucleon to derive the dark matter coupling to two nucleons. The trace anomaly of the QCD energy-momentum tensor implies~\cite{Shifman:1978zn,Belanger:2008sj}
\beqa
m_N\langle N | N\rangle = \langle N |   \sum_{1\leq i \leq n_f} m_i \overline{\psi}_i \psi_i ( 1+ \gamma) + \left(\frac{\beta^{n_f} }{2\,\alpha_s^2} \right)\,\alpha_s G^a_{\mu \nu} G^{a\, \mu \nu} | N \rangle \,,
\eeqa
where $\beta^{n_f} = -(11 - 2 n_f/3) \alpha^2_s/4\pi$ is the beta function at leading order, $n_f$ is the number of quarks, and $\gamma$ is the anomalous dimension of the quark field. So, at leading order in $\alpha_s$ and keeping only $n_f=3$ light quarks, we have
\beqa
 \langle N |\alpha_s G^a_{\mu \nu} G^{a\, \mu \nu} | N \rangle = \frac{8\pi}{9}\, m_N \left[\frac{1}{m_N}  \langle N |   m_u \overline{u}u + m_d \overline{d}d + m_s \overline{s}s | N  \rangle - 1 \right] \equiv  \frac{8\pi}{9}\,m_N \left[ (f_u + f_d + f_s) - 1 \right] \,.
\eeqa
Recent Lattice QCD updates of the calculation for the strange quark matrix element has $f_u+f_d+f_s = 0.085^{+0.022}_{-0.014}$~\cite{Junnarkar:2013ac}. In our numerical calculation, we will use $\langle N |\alpha_s G^a_{\mu \nu} G^{a\, \mu \nu} | N \rangle  \approx -2.56\,m_N$. The formula of the scattering cross section is then calculated to be
\beqa
\sigma^{\rm SI}_{X N} \,=\, \frac{\kappa^2 \, m_N^4}{4\pi\,\Lambda_1^4 \,(m_N + M_X)^2 } \,,
\eeqa
with $\kappa = \frac{2}{9} ( f_u + f_d + f_s - 1) \approx -0.20$. 

For the second operator, ${\cal O}^{\rm cRayleigh}_2$, we need to know the matrix element of $G\widetilde{G}$ inside a nucleon. This matrix element is related to the anomalous divergence of the iso-singlet axial current operator by
\beqa
\sum_{i=1,\cdots,n_f} \partial^\mu(\bar{\psi}_i \gamma_\mu \gamma_5 \psi_i) \, = \, \frac{n_f}{4\pi}{\alpha_s\,G^a_{\mu \nu} \widetilde{G}^{a\, \mu \nu} } \,+\,  \sum_{i=1,\cdots,n_f} 2i \,m_i \bar{\psi}_i \gamma_5 \psi_i \,.
\eeqa
In the large-$N_c$ and chiral limit and using the relation $\langle N | \bar{u}i\gamma_5 u + \bar{d}i\gamma_5 d + \bar{s}i\gamma_5 s | N \rangle = 0$, one finds~\cite{Cheng:2012qr,Hill:2014yka,Hill:2014yxa}
\beqa
\langle p | \frac{\alpha_s}{8\pi} G^a_{\mu \nu} \widetilde{G}^{a\, \mu \nu}  | p \rangle = 389\,\mbox{MeV}  \equiv \eta_p m_p \,, \quad 
\langle n | \frac{\alpha_s}{8\pi} G^a_{\mu \nu} \widetilde{G}^{a\, \mu \nu}  | n \rangle = -2\,\mbox{MeV} \equiv \eta_n m_n \,,
\eeqa
where the dimensionless parameters are $\eta_p \approx 0.41$ and $\eta_n \approx -0.0021$ (the instanton calculation in Ref.~\cite{Diakonov:1995qy} obtained dramatically different numbers). The large difference between $\eta_p$ and $\eta_n$ indicates a large isospin violation for pseudo-scalar coupling to nucleons. One can then use the matrix elements to translate ${\cal O}_2$ to the interaction between dark matter and nucleons, giving
\beqa
\frac{2i\,\eta_N\,m_N}{\Lambda_2^2}(XX - X^\dagger X^\dagger)\,\overline{N} i \gamma_5 N \,.
\eeqa
In the non-relativistic limit, one has $\overline{N} i \gamma_5 N \approx 2 i \vec{q} \cdot \vec{s}$ with $\vec{q}$ as the exchange momentum of the scattering process and $\vec{s}$ as the spin of a nucleon. So, for this interaction, we have both spin-dependent and momentum-suppressed scattering. The spin-dependent differential scattering cross section is
\beqa
\frac{d\sigma^{\rm SD}_{X\,N} }{d\cos{\theta}} = \frac{\eta_N^2\,m_N^2}{2\,\pi\,\Lambda_2^4} \frac{q^2}{(m_N + M_X)^2} \,.
\label{eq:scattering-O2}
\eeqa
For $M_X \sim 100$~GeV, a small cutoff scale of $\Lambda_2 \sim 100$~GeV, and a typical exchange momentum of $q\sim \mu_{X A}\,v \sim 100$~MeV, the spin-dependent scattering cross section of dark matter off a proton is $10^{-7}$~pb. This is far below the current direct detection experimental bound~\cite{Felizardo:2011uw,Behnke:2012ys}.

\subsection{Collider Constraints}
\label{sec:contact-collider}
For the two operators considered here, the universal signature at the LHC is that of a mono-jet plus missing transverse energy~\cite{Goodman:2010ku,Bai:2010hh}. For fermion dark matter coupling to two gluons, both the CMS~\cite{Khachatryan:2014rra} and ATLAS~\cite{ATLAS:2012zim} collaborations have imposed limits on the cutoffs of the effective operators. To estimate the constraints on our scalar dark matter case, we use \texttt{FeynRules}~\cite{Alloul:2013bka} to generate a model file for \texttt{MadGraph}~\cite{Alwall:2011uj}. We then use \texttt{Pythia}~\cite{Sjostrand:2007gs} to shower and hadronize the parton-level events. Finally, we use \texttt{PGS}~\cite{PGS} to cluster hadrons into jets and simulate detector effects. 

\begin{table}[ht!]
\vspace*{4mm}
\renewcommand{\arraystretch}{1.3}
\centerline{
\begin{tabular}{|c|c|c|}
\hline \hline
$M_X$~(GeV) &  $\Lambda_1$~(GeV)    &  $\Lambda_2$~(GeV)   \\  \hline
1    &     130    & 170    \\  \hline
10    &     120    & 180    \\  \hline
100  &  120    & 180   \\ \hline
200 & 110  & 160   \\ \hline
400  & 90    & 130 \\ 
 \hline \hline
 \end{tabular}
 \label{tab:monojet-constraint}
}
\caption{The collider constraints on the cutoff of the effective operators for different dark matter masses at 90\% CL. The mono-jet analysis with $E_{T}^{\rm miss} > 500$~GeV from the CMS collaboration in Ref.~\cite{Khachatryan:2014rra} has been used.
\label{tab:cutoff-constraint}} 
\end{table}

Following the same analysis procedure from the CMS collaboration in Ref.~\cite{Khachatryan:2014rra} at 8 TeV and with $19.7$~fb$^{-1}$ luminosity, we have found that imposing a cut on the missing transverse energy $E_{T}^{\rm miss} > 500$~GeV (or requiring less than 164 signal events) provides the strongest bounds for a wide range of dark matter masses. We show the constraints on the cutoffs of the effective operators in Table~\ref{tab:cutoff-constraint} for different dark matter masses. The constraints stay constant for light dark matter below 100 GeV and become weaker as one increases the dark matter mass beyond around 100 GeV. One can also see that for heavier dark matter beyond 200 GeV, the constraints become much weaker such that the cutoff is even below $M_X$, which indicates a breakdown of the perturbative description of the effective field theory.

\section{Simplified UV-completion Models}
\label{sec:simplified-model}
As already can be seen from Table~\ref{tab:cutoff-constraint}, the constraints on the cutoffs of the contact operators from the mono-jet searches are not that stringent. For a 100 GeV dark matter particle, the constrained cutoff is just comparable to the dark matter mass. This calls for UV-completed models to reduce the uncertainties from an effective field theory description. In this section, we consider several classes of models to illustrate that collider signatures beyond the mono-jet may provide a more sensitive probe of dark matter chromo-Rayleigh interactions. 

One of the simplest ways to UV-complete the two operators in Eqs.~(\ref{eq:operator1})(\ref{eq:operator2}) is to introduce a color-neutral scalar or pseudo-scalar, which can couple to two gluons either from the top quark or new heavy colored fermion loops. For the operator ${\cal O}^{\rm cRayleigh}_1$, one can simply introduce a Higgs-portal dark matter coupling like $XX^\dagger H H^\dagger$. With the effective coupling of the Higgs boson with two gluons in the SM, one generates the coupling of two dark matter particles to two gluons. For the second operator ${\cal O}^{\rm cRayleigh}_2$, the two-Higgs-doublet-portal dark matter models have a pseudo-scalar as a mediator to generate the dark matter chromo-Rayleigh interactions (see Refs.~\cite{Bai:2012nv,Buckley:2014fba,Berlin:2015wwa,Haisch:2015ioa}). The existing search strategy for the Higgs-portal or pseudo-scalar-portal dark matter should cover this class of UV-completion models~\cite{Chatrchyan:2014tja}. Therefore, we do not discuss this case in great detail. Instead, we consider two other ways to UV-complete the two operators and point out more interesting collider signatures. 

\subsection{QCD-charged Particle Mediation for ${\cal O}^{\rm cRayleigh}_1$}
\label{sec:colored-O1}
In this subsection, we examine a class of models with additional QCD-charged scalars to generate the effective chromo-Rayleigh interaction of ${\cal O}^{\rm cRayleigh}_1$. For specificity's sake, we introduce a real color-octet and electroweak-singlet scalar,~\footnote{A similar analysis can be performed for other QCD representations.} $G_H^a$ with $a=1, \cdots, 8$ and a mass of $M_{G_H}$, which, for example, appears in the Renormalizable Coloron Model (ReCoM)~\cite{Bai:2010dj,Chivukula:2013xka}. At the renormalizable level, one has the following interaction coupling two $G_H$'s to two dark matter fields
\beqa
{\cal L} \supset - \frac{\lambda}{2} \, G_H^a\, G_H^a\, X^\dagger X \,.
\eeqa
\begin{figure}[t]
\begin{center}
\includegraphics[width=0.8\textwidth]{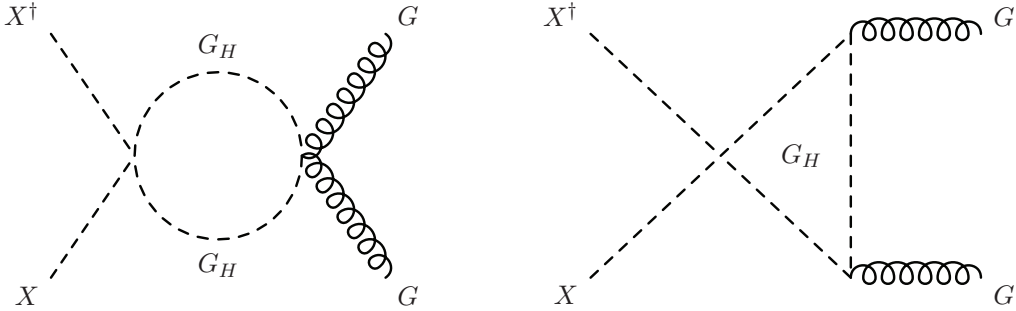} 
\caption{Representative loop diagrams for generating ${\cal O}^{\rm cRayleigh}_1$ via colored scalar loops.}
\label{fig:feynman-bubble}
\end{center}
\end{figure}

At one-loop level, the effective operator can be generated through the diagrams in Fig.~\ref{fig:feynman-bubble} and has the following calculated form
\beqa
F(\tau)\, \frac{\lambda}{8\,M_{G_H}^2} \, \frac{\alpha_s}{4\pi}X^\dagger X \,G^a_{\mu\nu} G^{a\, \mu\nu} \,,
\eeqa
where the form factor $F(\tau)$, with $\tau = (p_1+p_2)^2/(4M^2_{G_H})$, is one in the limit $\tau \rightarrow 0$. Here, $p_1$ and $p_2$ are the gluon momenta pointing to the vertex. In the limit that $M_{G_H} \gg M_X$ and with small gluon momenta much below $|p_1+p_2|$, one can match the colored particle mass to the effective operator cutoff defined in Eq.~(\ref{eq:operator1}) to obtain
\beqa
\Lambda_1^2 = \frac{8}{\lambda}\, M^2_{G_H} \,.
\label{eq:cutoff-convert}
\eeqa

For this class of UV-complete models, one can also work out the collider constraints.  Depending on whether the color-octet is odd or even under the dark matter ${\cal Z}_2$, one has different decay channels for $G_H^a$. If it is ${\cal Z}_2$-even, the following renormalizable $G_H$ cubic self-interaction can make $G_H$ unstable
\beqa
\mu_G \, d_{abc}\, G_H^a G_H^b G_H^c \,,
\eeqa
with $d_{abc}$ the totally-symmetric $SU(3)_{\rm QCD}$ tensor. This introduces the decay of $G_H$ into two gluons at one loop with a width of~\cite{Gresham:2007ri,Bai:2010dj}
\beqa
\Gamma(G_H \rightarrow gg) = \frac{15\,\alpha_s^2\,\mu_G^2}{128\,\pi^3\,M_{G_H}} \left( \frac{\pi^2}{9} - 1 \right)^2 \,.
\eeqa
$\mathcal{Z}_2$-even $G_H$ can be pair-produced at the LHC from their QCD interactions, which then decay into paired dijets. Following the same reinterpretation of the experimental data as in Ref.~\cite{Bai:2013xla}, we find that  the current searches for paired dijet resonances at the 8 TeV LHC have set a constraint on its mass of~\cite{Khachatryan:2014lpa}
\beqa
M_{G_H} \gtrsim 520~\mbox{GeV}\,, \qquad \qquad  \mbox{for }{\cal Z}_2\mbox{-even }  G_H \,,
\eeqa
where we have only included the QCD productions. 

If $G_H$ is $\mathcal{Z}_2$-odd, the dark matter particle $X$ has to appear in the $G_H$ decay products. The operators mediating $G_H$ decaying to $X$ first occur at the dimension six level and contain the following two parity-conserving operators
\beqa
\frac{D_\mu G_H^a \partial_\nu X \, G^{a\,\mu\nu}}{\Lambda_1^{\prime\,2}} \,+\, h.c.\,,   \qquad \qquad \qquad
\frac{G_H^a X \widetilde{H} \overline{Q}_L t^a t_R} {\Lambda_1^{\prime\prime\,2}} \,+\, h.c.  \,.
\label{eq:GH-decay-operator}
\eeqa
Here, $t^a$ with $a=1,2, \cdots,8$ are the $SU(3)_{\rm QCD}$ generators. For the second operator, we have only included the top quark by assuming that the coupling is proportional to the quark mass. Using the equation of motion, $D_\mu G^{a\,\mu\nu} = - g_s \,\bar{q}\gamma^\nu t^a q$, the first decay-inducing operator becomes
\beqa
\frac{g_s}{\Lambda_1^{\prime\,2}}\,G_H^a \partial_\nu X\, \bar{q}\gamma^\nu t^a q \,.
\eeqa
As a result, the decay channels of the colored state $G_H$ are mainly $G_H \rightarrow X \bar{q} q$. For each flavor, the three-body decay width is calculated to be
\beqa
&&\hspace{-0.5cm}\Gamma(G_H \rightarrow X \bar{q} q) =   \nonumber \\
&&\hspace{-0.5cm}\frac{g_s^2}{3\cdot 2^9\,\pi^3 \Lambda_1^{\prime\, 4}} 
\int_{4 m_q^2}^{(M_{G_H} - M_X)^2} ds \left ( 1 + \frac{2 m_q^2}{s} \right ) \left ( 1 - \frac{4 m_q^2}{s} \right )^{1/2}    \, \left [ \frac{M_X^4 + (M_{G_H}^2 - s)^2 - 2 M_X^2 (M_{G_H}^2 + s)}{M_{G_H}^2} \right ]^{3/2} \,.
\eeqa
Choosing $g_s = 1.1$, $M_{G_H}=500$~GeV, $M_X = 10$~GeV, $\Lambda_1^\prime = 1$~TeV, and neglecting the quark mass, $\Gamma = 0.0002$~GeV for each flavor and $G_H$ can decay promptly for collider studies.

From the above assumptions, the collider signature for pair-produced $\mathcal{Z}_2$-odd $G_H$ contains $4j + E_T^{\rm miss}$, $t\bar{t}+2j + E_T^{\rm miss}$ and $2t+2\bar{t}+E_T^{\rm miss}$. Generically, the top-quark rich final state can be easily searched for at the LHC. So, we concentrate on the first operator in Eq.~(\ref{eq:GH-decay-operator}) and derivate a more conservative bound based mainly on the $4j + E_T^{\rm miss}$ final state. Following the analysis in Ref.~\cite{Chatrchyan:2014lfa} at the 8 TeV LHC with 19.5 fb$^{-1}$, we find that the set of cuts with $300< \slsh{H}_T < 450$~GeV, $800<H_T < 1000$~GeV, $3\le N_{\rm jets} \le 5$ with $p_T(j) > 50$~GeV and $|\eta| < 2.5$ provide the best constraint. For $M_{G_H}=600$~GeV and $M_X = 10$~GeV, we have $\mbox{Br}(G_H \rightarrow 2j + X) = 97.5\%$ and $\mbox{Br}(G_H \rightarrow t\bar t + X) = 2.5\%$. The production cross section at the 8 TeV LHC after this set of cuts is approximately $1.9\,\mbox{fb}\times K$ with the $K$-factor as 1.8~\cite{GoncalvesNetto:2012nt}. This amounts to a total of 65.4 events at 19.5 fb$^{-1}$, which is very close to the allowed number of signal events (65.6 from Ref.~\cite{Chatrchyan:2014lfa}) at 90\% C.L.  So, for a large mass splitting between the color-octet state $G_H$ and the dark matter $X$, the current LHC bound is
\beqa
M_{G_H} \gtrsim 600~\mbox{GeV}\,, \qquad \qquad  \mbox{for }{\cal Z}_2\mbox{-odd }  G_H \,.
\eeqa

Independent on how $G_H$ decays, we have set a constraint on the $G_H$ mass to be above 500-600 GeV. Using the relation in Eq.~(\ref{eq:cutoff-convert}), this constraint translates to a limit on the effective cutoff scale of $\Lambda_1 \gtrsim 1.5-1.7$~TeV for $\lambda=1$, which is far more stringent than the limits in Table~\ref{tab:cutoff-constraint} set from the mono-jet search. To compare with the limits from direct detection experiments, we show the reinterpreted collider limits using $\Lambda_1 > 1.7$~TeV for the first chromo-Rayleigh interaction in the left panel of Fig.~\ref{fig:contact-scattering}. The left panel of Fig.~\ref{fig:contact-scattering} shows a comparison between the limits on the scattering cross section $\sigma^{\rm SI}_{X N}$ obtained from direct detection, monojet, and $4j + E_T^{\rm miss}$ searches.

\subsection{QCD-charged Particle Mediation for ${\cal O}^{\rm cRayleigh}_2$}
\label{sec:colored-O2}
To UV-complete the operator ${\cal O}^{\rm cRayleigh}_2$, we introduce the following $CP$-conserving Lagrangian
\beqa
{\cal L} \supset - y_1 \left( X + X^\dagger\right) (\overline{\psi}_1 \psi_2 + \overline{\psi}_2 \psi_1) 
-   \left(X - X^\dagger \right)  (y_2\,\overline{\psi}_1\gamma_5 \psi_2 + y_2\,\overline{\psi}_2 \gamma_5 \psi_1) \,.
\eeqa
Here, $\psi_1$ and $\psi_2$ are chosen to be QCD triplets and their electroweak quantum numbers will be discussed and specified later. To conserve the dark ${\cal Z}_2$ symmetry, one of  the fermion triplets must be $\mathcal{Z}_2$-odd while the other is $\mathcal{Z}_2$-even. Using the freedom of field redefinitions of $\psi_1$ and $\psi_2$, one can keep the first coupling, $y_1$, to be a real number while the second coupling, $y_2$, may in general be complex. We choose $y_2$ to be a real number to satisfy $CP$ symmetry.  In terms of the components, $X = (X_R + i\, X_I)/\sqrt{2}$, we have the interactions
\beqa
{\cal L} \supset - \sqrt{2}\,y_1\,X_R\, (\overline{\psi}_1 \psi_2 + \overline{\psi}_2 \psi_1) 
-\sqrt{2}\,i\,X_I\, (y_2\,\overline{\psi}_1\gamma_5 \psi_2 + y_2\,\overline{\psi}_2 \gamma_5 \psi_1) \,.
\label{eq:XI-lagrangian}
\eeqa
%
To preserve $C$ and $P$, one has $X_R$ to be $C$-even and $P$-even, and $X_I$ to be $C$-even and $P$-odd. In terms of $X_R$ and $X_I$, we have the effective operator
\beqa
{\cal O}^{\rm cRayleigh}_2 = - \frac{\alpha_s}{2 \pi \Lambda_2^2} X_R X_I G^a_{\mu \nu} \widetilde{G}^{a\, \mu \nu}  \,,
\eeqa
which conserves both $C$ and $P$. 

\begin{figure}[th!]
\begin{center}
\includegraphics[width=0.48\textwidth]{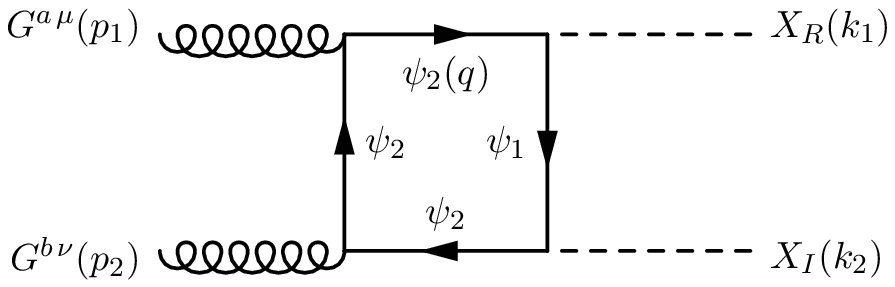} \hspace{3mm}
\includegraphics[width=0.48\textwidth]{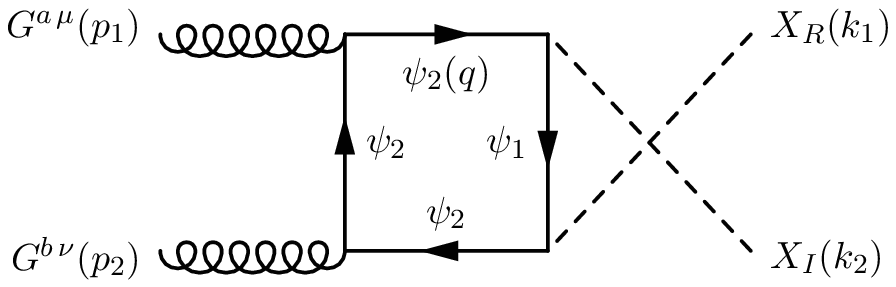} 
\caption{Representative loop diagrams for generating ${\cal O}^{\rm cRayleigh}_2$ via colored fermion loop.}
\label{fig:box-diagram-example}
\end{center}
\end{figure}

At one-loop level, one has the box diagrams in Fig.~\ref{fig:box-diagram-example} to generate the effective operator ${\cal O}^{\rm cRayleigh}_2$. 
In the heavy particle limit with $m_{\psi_1}, m_{\psi_2} \gg M_{X_R}, M_{X_I}$, we have the matching condition (see Appendix~\ref{appendix} for a more detailed calculation)
\beqa
\Lambda^2_2 = \frac{ 2\,m_{\psi_1} m_{\psi_2} }{|y_1 y_2|}  \,.
\label{eq:cutoff-convert-2}
\eeqa
For this specific UV-completion model, the first chromo-Rayleigh interaction can also be generated, since it does not break any discrete symmetries.  

If $\mbox{Im}(y_1 y_2) \neq 0$, the Lagrangian in Eq.~(\ref{eq:XI-lagrangian}) is $P$-conserving but $C$-breaking, so $CP$ is also broken.~\footnote{For a complex $y_2$ with $CP$-violating interactions, one may wonder about generating the Weinberg operator~\cite{Weinberg:1989dx}, $f_{abc}G^{a}_{\mu\rho}G^{b\,\rho}_\nu\widetilde{G}^{c\,\mu\nu}$. We note that the $P$ conservation forbids the generation of the effective Weinberg operator.} The lowest dimensional effective operator that is $C$-odd and $P$-even is at dimension-10, for instance $X_RX_I\,d_{abc}(D_\mu G^a_{\alpha\beta})(D_\nu G^{b\,\nu \alpha}) \widetilde{G}^{c\,\beta \mu}$, further suppressed by powers of the cutoff scale. If parity was broken, lower-dimensional operators like $X_R X_I G^a_{\mu\nu}G^{a\,\mu\nu}$ could be generated with a stringent bound from the neutron electric dipole moment.

As mentioned before, in order to conserve the dark ${\cal Z}_2$ symmetry the two fields, $\psi_1$ and $\psi_2$, should have opposite dark parity. Fixing the mass relation $m_{\psi_2} > m_{\psi_1}$, we have two cases. The case A has ${\cal Z}_2$-even $\psi_1$ and ${\cal Z}_2$-odd $\psi_2$ and the case B has ${\cal Z}_2$-even $\psi_2$ and ${\cal Z}_2$-odd $\psi_1$.  For both cases, we introduce the following dimension-5 operator to mediate the ${\cal Z}_2$-even particle decaying into two jets,
\beqa
\frac{g_s^2}{4\pi\,\Lambda_2^\prime} \bar{\psi}_{i\,L} \sigma^{\mu\nu} t^a u_R\,\widetilde{G}_a^{\mu\nu} \,.
\label{eq:psi-decay-operator}
\eeqa
Here, $i=1 (2)$ for case A(B); the electroweak quantum numbers of $\psi_i$ are then the same as $u_R$ (or $d_R$). As a result of this decay mode, the unstable particle $\psi_i$ behaves as a dijet resonance at colliders. One could also consider other potential decay channels by introducing the electroweak dipole moment operator, which has a clearer signature at colliders. 

\begin{figure}[th!]
\begin{center}
\includegraphics[width=0.48\textwidth]{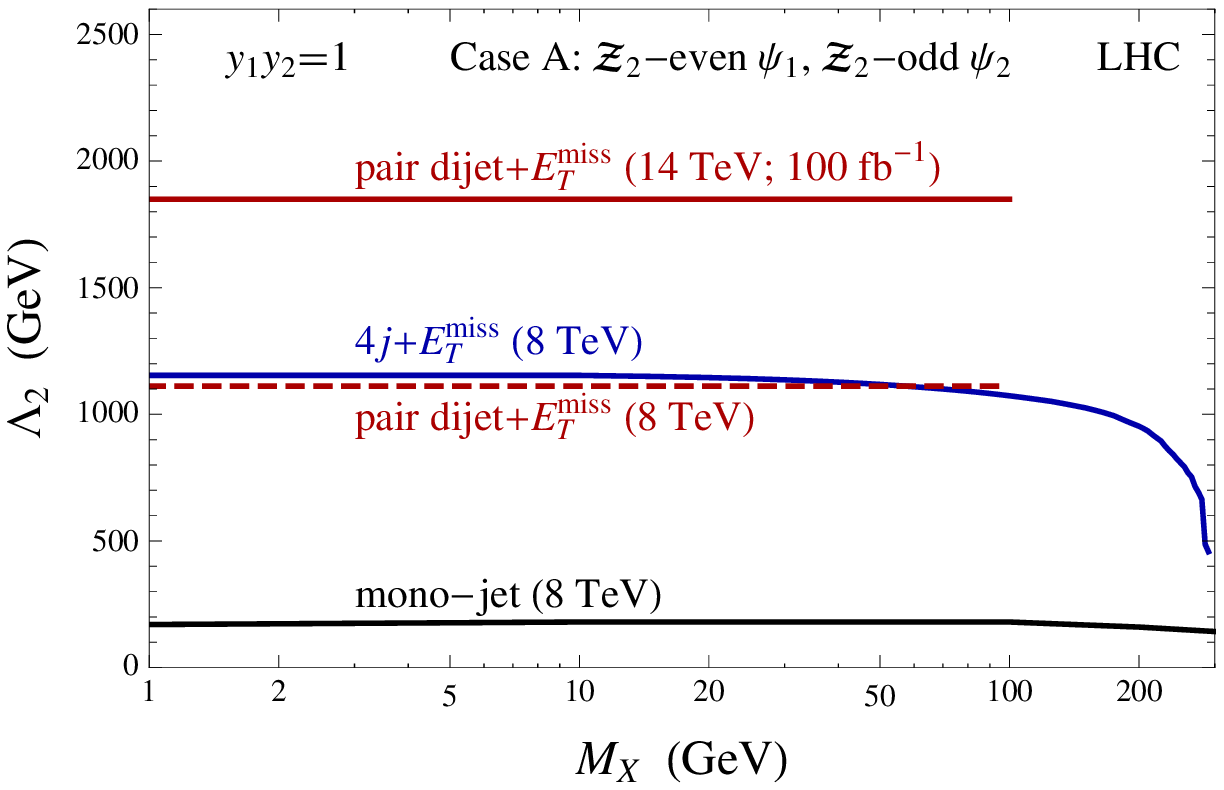} \hspace{3mm}
\includegraphics[width=0.48\textwidth]{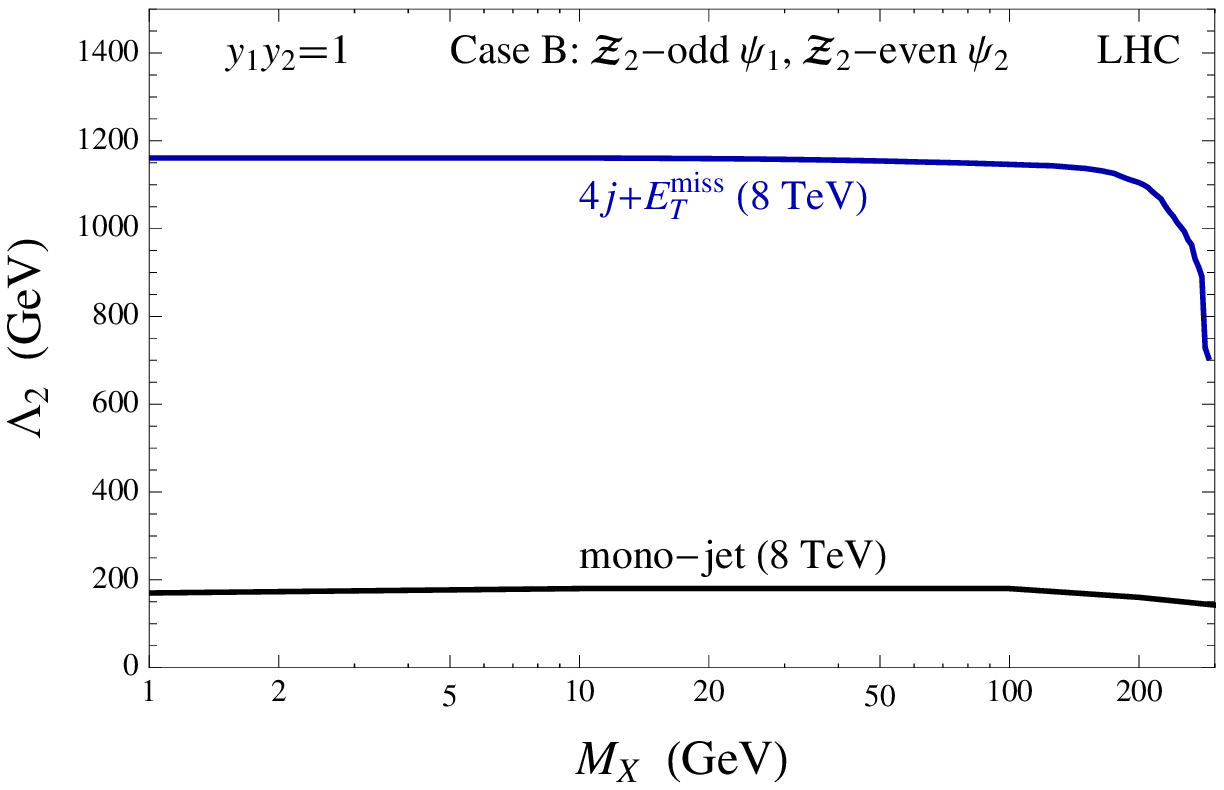} 
\caption{{\bf Left panel:} constraints on the cutoff of the effective operator ${\cal O}_2^{\rm cRayleigh}$ for case A with ${\cal Z}_2$-even $\psi_1$ and ${\cal Z}_2$-odd $\psi_2$.
 {\bf Right panel:} constraints for case B with ${\cal Z}_2$-odd $\psi_1$ and ${\cal Z}_2$-even $\psi_2$.}
\label{fig:O2-cutoff-constraint}
\end{center}
\end{figure}

For case A with a ${\cal Z}_2$-even $\psi_1$, the collider signature of $\psi_1$ is pair-produced dijet resonances. The searches at the 8 TeV LHC have set a constraint on its mass~\cite{Khachatryan:2014lpa} as $m_{\psi_1} \gtrsim  500~\mbox{GeV}$. Although the signature is interesting by itself, the discovery of this dijet resonance can not prove that dark matter has been produced at colliders. For the ${\cal Z}_2$-odd $\psi_2$ particle, ithe decay channel is $\psi_2 \rightarrow X + \psi_1 \rightarrow X + 2j$. Depending on the mass splitting of $\psi_1$ and $\psi_2$, one has $\psi_1$ to be off-shell for $m_{\psi_2} - m_{\psi_1} < M_X$ and on-shell for $m_{\psi_2} - m_{\psi_1} > M_X$. For the off-shell intermediate $\psi_1$ case, this signature is very similar to the SUSY squark searches with a heavy gluino~\cite{Chatrchyan:2014lfa}, except with a larger production cross section than a single flavor squark. For dark matter mass $M_X \lesssim 100$~GeV, the constraint on the $\psi_2$ mass is $m_{\psi_2} \gtrsim 850$~GeV, assuming that the signal acceptance is similar to the squark one~\cite{Chatrchyan:2014lfa}. For a heavy dark matter mass close to 290~GeV, the constraint becomes weaker and is $m_{\psi_2} \gtrsim 500$~GeV. Using the conversion formula in Eq.~(\ref{eq:cutoff-convert-2}), we show the constraints from $4j+E_T^{\rm miss}$ on the effective cutoff for different dark matter masses in the left panel of Fig.~\ref{fig:O2-cutoff-constraint}, by setting $m_{\psi_1}=m_{\psi_2}-M_X$ and $y_1y_2=1$. 

For the on-shell $\psi_1$ case, the signature is more interesting with a pair of dijet resonances plus a large $E_T^{\rm miss}$. The relevant production Feynman diagram is shown in the left panel of Fig.~\ref{fig:feynman-psi2}. We perform a detailed collider study for this interesting signature in Section~\ref{sec:collider-pair-dijet-met}. As a comparison, in the left panel of Fig.~\ref{fig:O2-cutoff-constraint} we show the interpreted constrains for $M_X \lesssim 100$~GeV on the matched cutoff from Eq.~(\ref{eq:cutoff-convert-2}) for both the 8 TeV LHC with 19.5 fb$^{-1}$ and the 14 TeV LHC with 100 fb$^{-1}$. One can see a clear improvement on constraining this effective chromo-Rayleigh interaction from a dedicated search beyond the simple mono-jet signature. 

For case B with ${\cal Z}_2$-odd $\psi_1$ and ${\cal Z}_2$-even $\psi_2$, the searches for pair-produced dijet resonances provide a constraint of $m_{\psi_2} \gtrsim 500$~GeV. The decay of $\psi_1$ needs to go through an off-shell $\psi_2$ via $\psi_1\rightarrow X + \psi_2^*  \rightarrow X+ 2j$. Applying the SUSY squark searches with a heavy gluino~\cite{Chatrchyan:2014lfa}, the constraint on the $\psi_1$ mass is $m_{\psi_2}\ge m_{\psi_1} \gtrsim 850$~GeV for $M_X \lesssim 100$~GeV and $m_{\psi_1} \gtrsim 500$~GeV for $M_X$ close to 290~GeV. In the right panel of Fig.~\ref{fig:O2-cutoff-constraint}, we show the reinterpreted constraints on the effective cutoff for $y_1y_2=1$.

\subsection{Pair-produced Dijet Resonances plus $E_T^{\rm miss}$}
\label{sec:collider-pair-dijet-met}
\begin{figure}[th!]
\begin{center}
\includegraphics[width=0.44\textwidth]{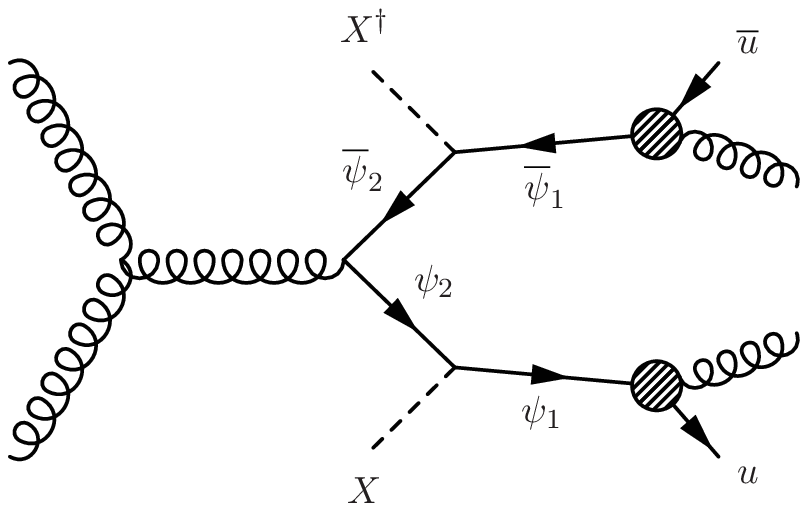} \hspace{3mm}
\includegraphics[width=0.46\textwidth]{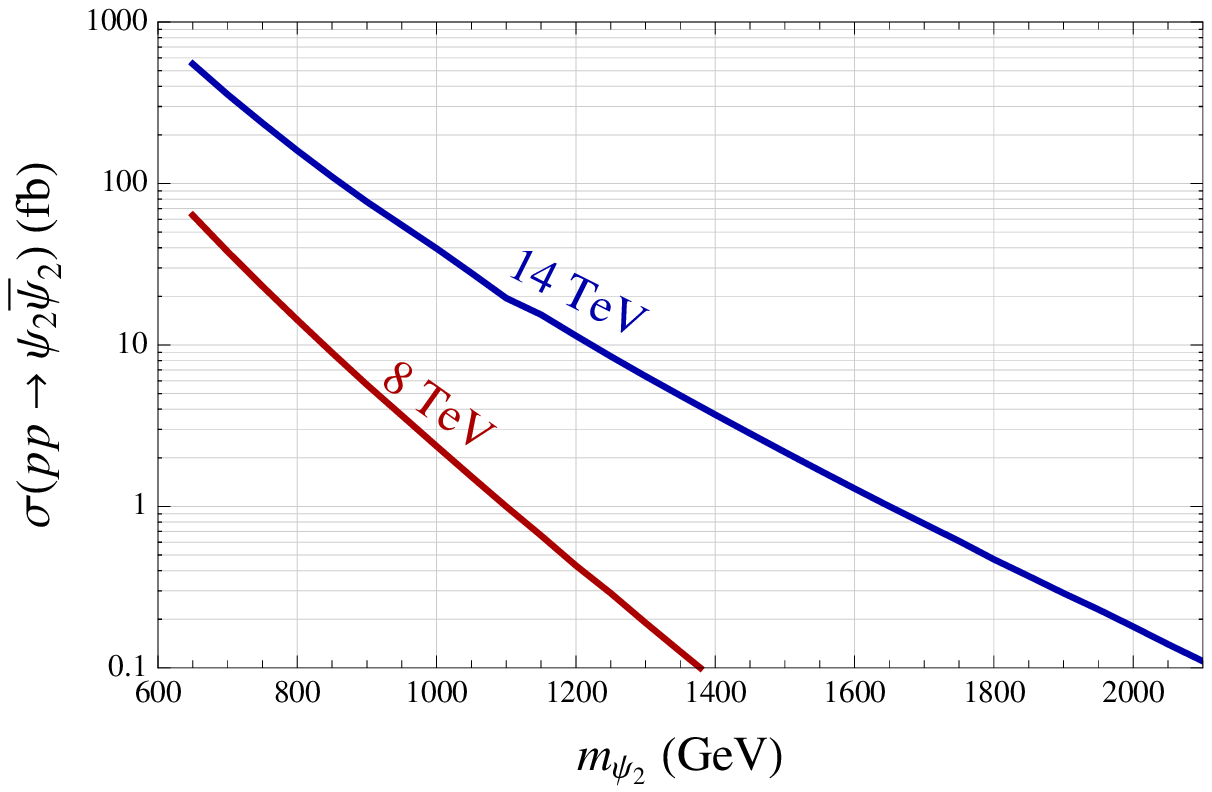} 
\caption{{\bf Left panel:} the representive Feynman diagram for the process of $pp \rightarrow \psi_2 \bar{\psi}_2 \rightarrow X + \psi_1 + X^\dagger + {\bar \psi}_1 \rightarrow X X^\dagger+ 4j$. {\bf Right panel:} the tree-level production cross sections of $pp \rightarrow \psi_2 \bar{\psi}_2$. }
\label{fig:feynman-psi2}
\end{center}
\end{figure}

In this section, we perform a detailed collider study for the process of $pp \rightarrow \psi_2 \bar{\psi}_2 \rightarrow X + \psi_1 + X^\dagger + {\bar \psi}_1 \rightarrow X X^\dagger+ 4j$. The collider signature is pair-produced dijet resonances plus missing transverse energy. In the left panel of Fig.~\ref{fig:feynman-psi2}, we show the Feynman diagram for this process.
The QCD production cross sections for $\psi_2 \bar{\psi}_2$ at the LHC are the same as vector-like $t^\prime\bar{t^\prime}$~\cite{Chatrchyan:2013uxa}. We show the tree-level production cross sections, calculated using \texttt{MadGraph}~\cite{Alwall:2011uj}, in the right panel of Fig.~\ref{fig:feynman-psi2}. In the following analysis, we will ignore the signal $K$-factor because we will use tree-level cross sections for backgrounds.~\footnote{One might expect the overall significance in the later analysis to be increased by a factor of $\sim 1.2 - 1.3$ due to the inclusion of NLO effects.}

The main SM background comes from $Z/W^\pm+n$ jets with $Z\rightarrow \nu\bar{\nu}$ and leptonic decays of $W^\pm$. After comparing the results from matching parton showers and matrix elements~\cite{Hoche:2006ph}, we have found that the background of parton-level $Z/W^\pm+4$ jets with $p_T(j)>120$~GeV provides a good estimation of total $Z/W^\pm+n$ jets background for the 8 TeV LHC. Therefore, we use $Z/W^\pm+4$ jets as an approximation to save simulation time. There also exist additional, sub-dominant semi-leptonic $t\bar{t}$ backgrounds, which will be kept in our analysis. 

\begin{figure}[th!]
\begin{center}
\includegraphics[width=0.44\textwidth]{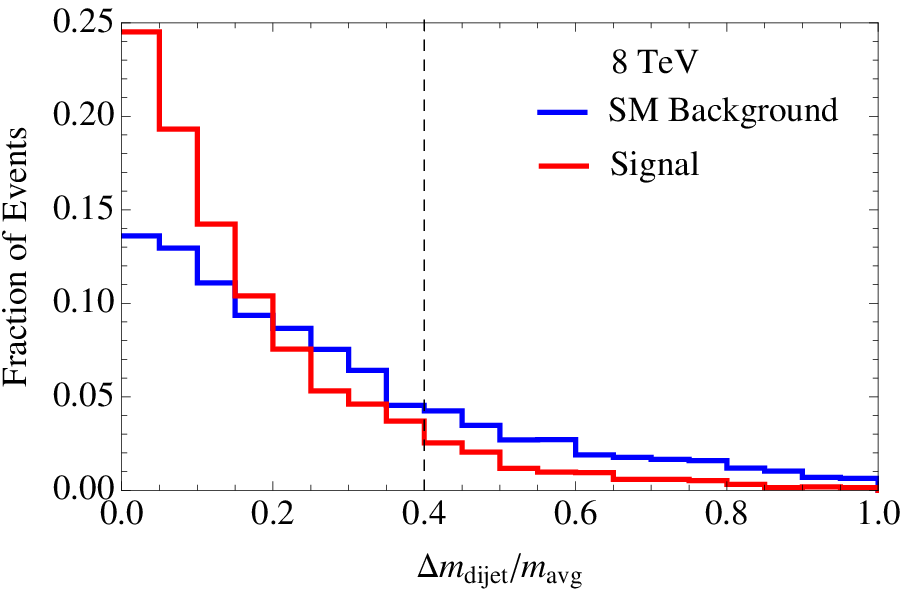} \hspace{3mm}
\includegraphics[width=0.45\textwidth]{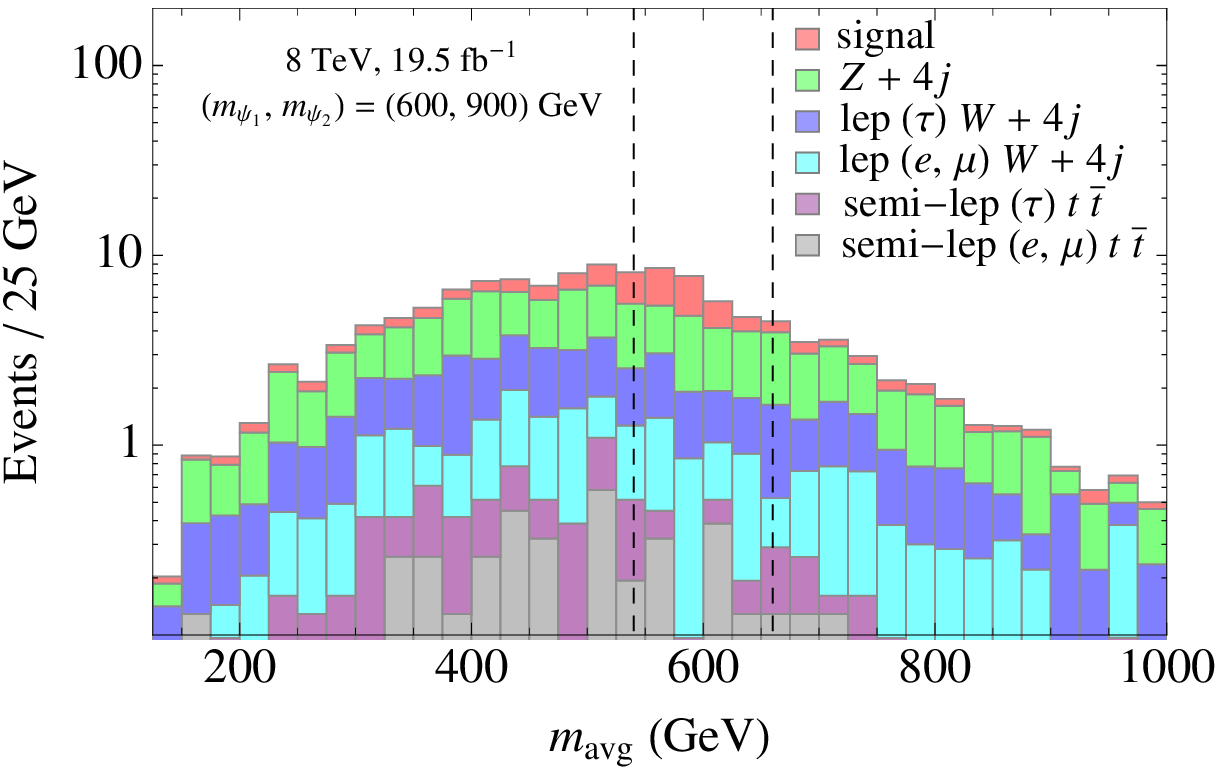} 
\caption{{\bf Left panel:} the distributions of the fraction of events as a function of $\Delta m_{\rm dijet} / m_{\rm avg}$ for the signal of $(m_{\psi_1}, m_{\psi_2}) = (600, 900)$~GeV and the summed background. {\bf Right panel:} the averaged dijet invariant mass distributions for the signal and various backgrounds, after the cut of $\Delta m_{\rm dijet} / m_{\rm avg} < 0.4$.}
\label{fig:mass-diff}
\end{center}
\end{figure}

\begin{figure}[th!]
\begin{center}
\includegraphics[width=0.46\textwidth]{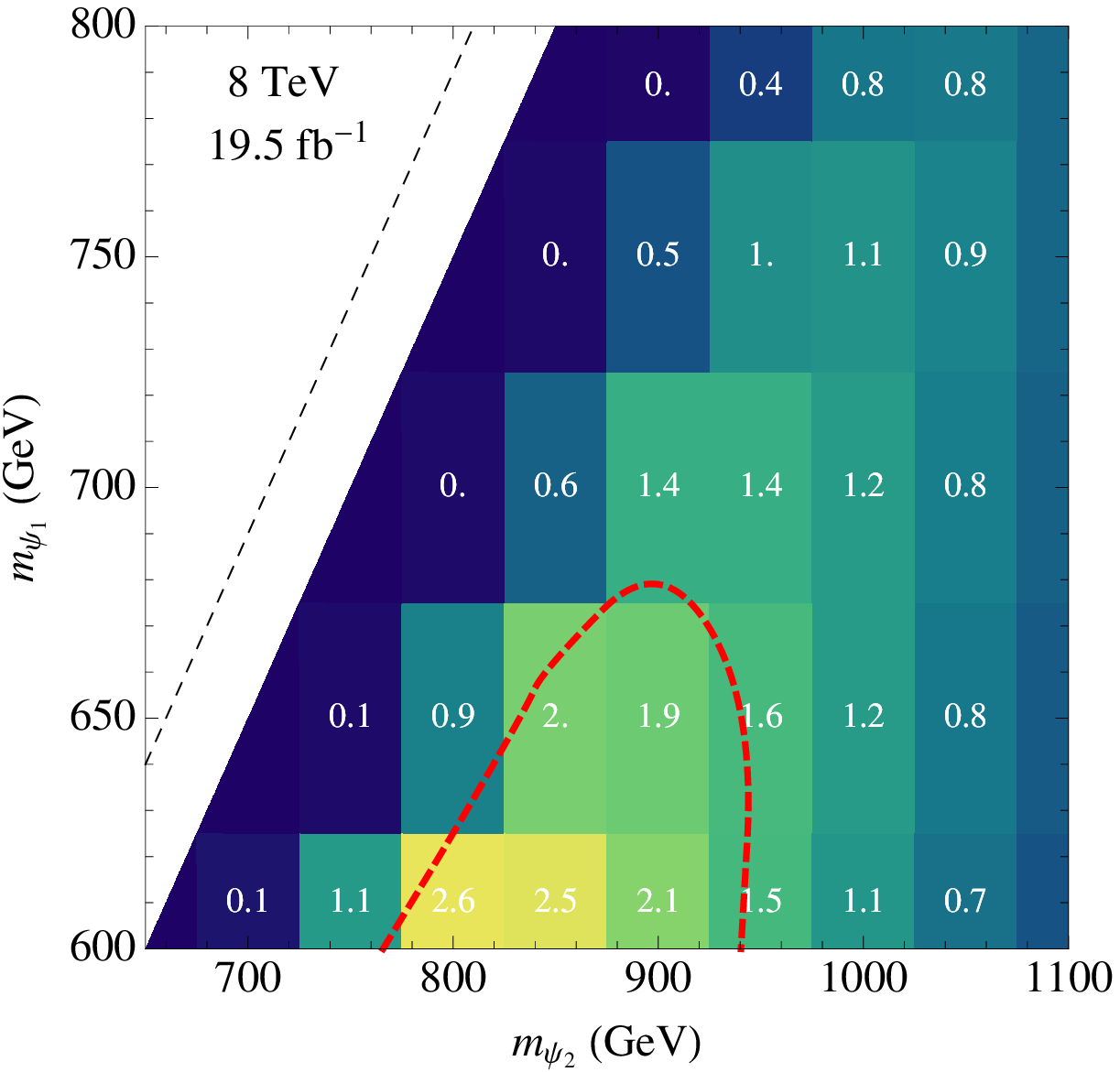}  \hspace{3mm}
\includegraphics[width=0.46\textwidth]{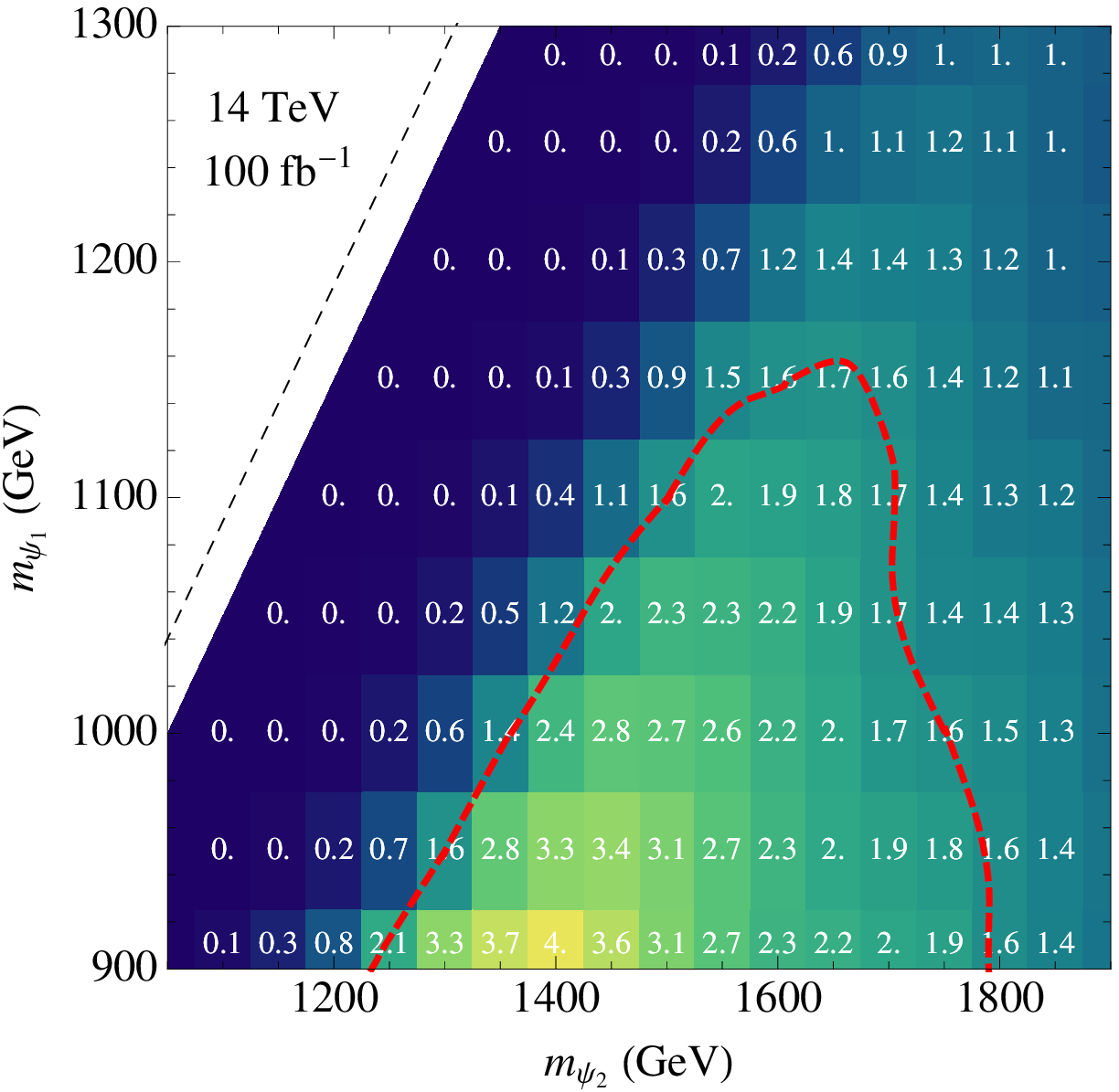}
\caption{{\bf Left panel:} the signal significance $S/\sqrt{B}$ for different model points with the cuts optimized for $(m_{\psi_1}, m_{\psi_2}) = (600, 900)$~GeV and $M_X=10$~GeV at the 8 TeV LHC with 19.5 fb$^{-1}$. The red dashed line is the extrapolated 90\% C.L. exclusion curve. The black dashed line has $m_{\psi_2} = m_{\psi_1} + M_X$. {\bf Right panel:} the same as the left panel but for 14 TeV with 100 fb$^{-1}$ and the set of cuts  for the combined benchmark points $(m_{\psi_1}, m_{\psi_2}) = (900, 1800)$~GeV and $(900,1500)$~GeV with $M_X=10$~GeV.}
\label{fig:significance-map}
\end{center}
\end{figure}

Starting from the 8 TeV LHC with 19.5 fb$^{-1}$, we choose the model point $(m_{\psi_1}, m_{\psi_2}) = (600, 900)$~GeV as a benchmark to optimize our cuts on kinematic variables. We choose the basic cuts on the jet and missing transverse momenta to be $p_T(j_i) > 140$~GeV and $E_T^{\rm miss} > 275$~GeV and required at least four jets satisfying the jet $p_T$ cut in the final state. To reduce the $t\bar{t}$ and $W^\pm+$jets backgrounds, we also veto events containing a lepton with $p_T(\ell) > 20$~GeV. Since our signal has a pair of dijet resonances with the same mass, we choose the combination among three possible dijet pairs with the smallest dijet invariant mass difference, $\Delta m_{\rm dijet}$. To further reduce the SM background, we show the event fraction histogram distribution in terms of the variable $\Delta m_{\rm dijet} / m_{\rm avg}$ in the left panel of Fig.~\ref{fig:mass-diff}. Signal events prefer a smaller value of dijet invariant mass difference than background events. As a result, we impose a cut on this variable with $\Delta m_{\rm dijet} / m_{\rm avg} < 0.4$ to further increase the discovery sensitivity. In the right panel of Fig.~\ref{fig:mass-diff}, we show the averaged dijet invariant mass distribution for the signal and backgrounds. Since the signal events mainly distribute around the $\psi_1$ particle mass, we also impose one additional cut with $|m_{\rm avg}-m_{\psi_1}|<0.1\, m_{\psi_1}$ to further improve the sensitivity. With the above cuts, we show the values of $S/\sqrt{B}$ for different mass points in the left panel of Fig.~\ref{fig:significance-map}. At 90\% C.L., our simulated results show that the model point of $(m_{\psi_1}, m_{\psi_2}) = (650, 950)$~GeV can be covered, which corresponds to $\Lambda_2 \gtrsim 1.1$~TeV. Once again we see a dramatic improvement relative to the standard mono-jet signature.

At the 14 TeV LHC with 100 fb$^{-1}$, we choose two benchmark points with $M_X = 10$~GeV and $(m_{\psi_1}, m_{\psi_2}) = (900, 1800) \;[(900, 1500)]$~GeV. To optimize the discovery sensitivity, we choose the following set of cuts: at least four jets with $p_T(j) > 200$~GeV; $E_T^{\rm miss} > 900 \;[600]$~GeV; $\Delta m_{\rm dijet} / m_{\rm avg} < 0.4 \;[0.3]$; $|m_{\rm avg}-m_{\psi_1}|<0.1\, m_{\psi_1}$. The signal significances, $S/\sqrt{B}$, for different model points are shown in the right panel of Fig.~\ref{fig:significance-map}, taking the larger $S / \sqrt{B}$ value from between the two benchmark points. We show the combined 90\% C.L. exclusion contour in the red dashed line, which has the lower right region covered by the set of cuts for $(900, 1800)$~GeV and the upper right region by $(900, 1500)$~GeV. At 90\% C.L., one can see that the model point of $(m_{\psi_1}, m_{\psi_2}) = (950, 1800)$~GeV can be excluded, which corresponds to $\Lambda_2 \gtrsim 1.8$~TeV.

\section{Comparison to the Direct Detection Limits}
\label{sec:compare}
\begin{figure}[th!]
\begin{center}
\includegraphics[width=0.48\textwidth]{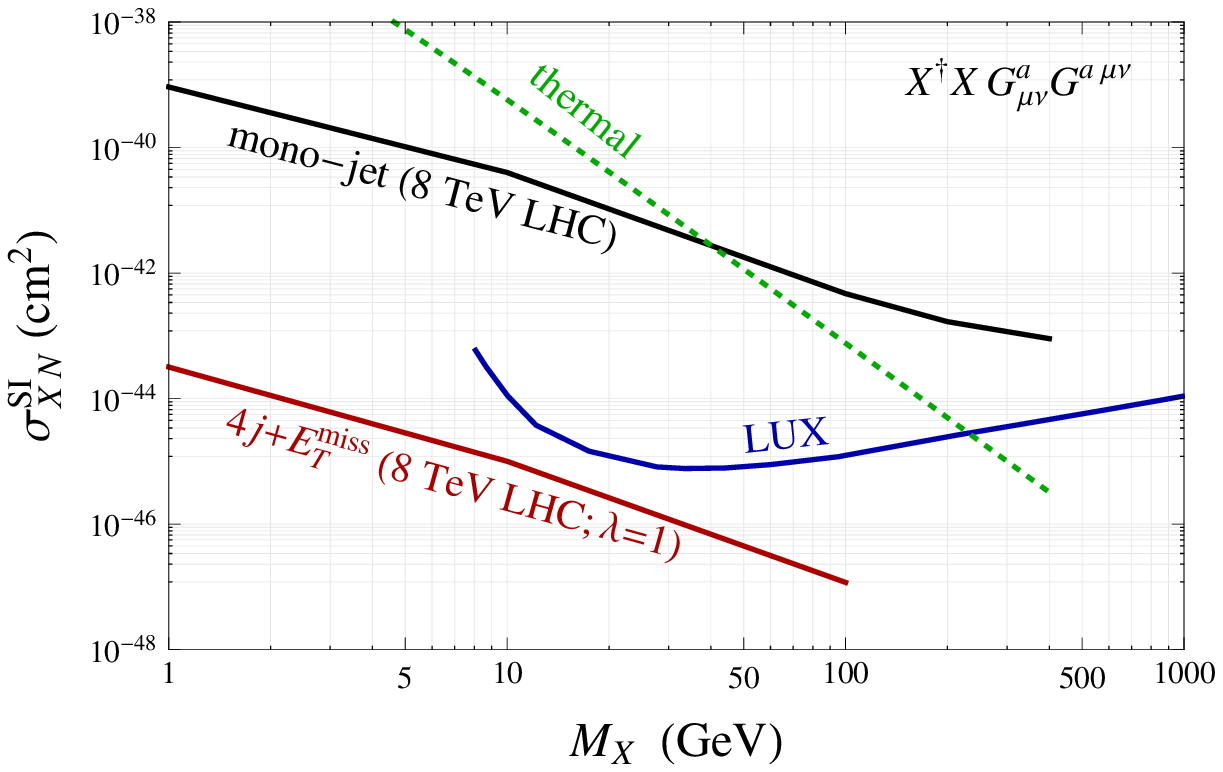} 
\hspace{3mm}
\includegraphics[width=0.48\textwidth]{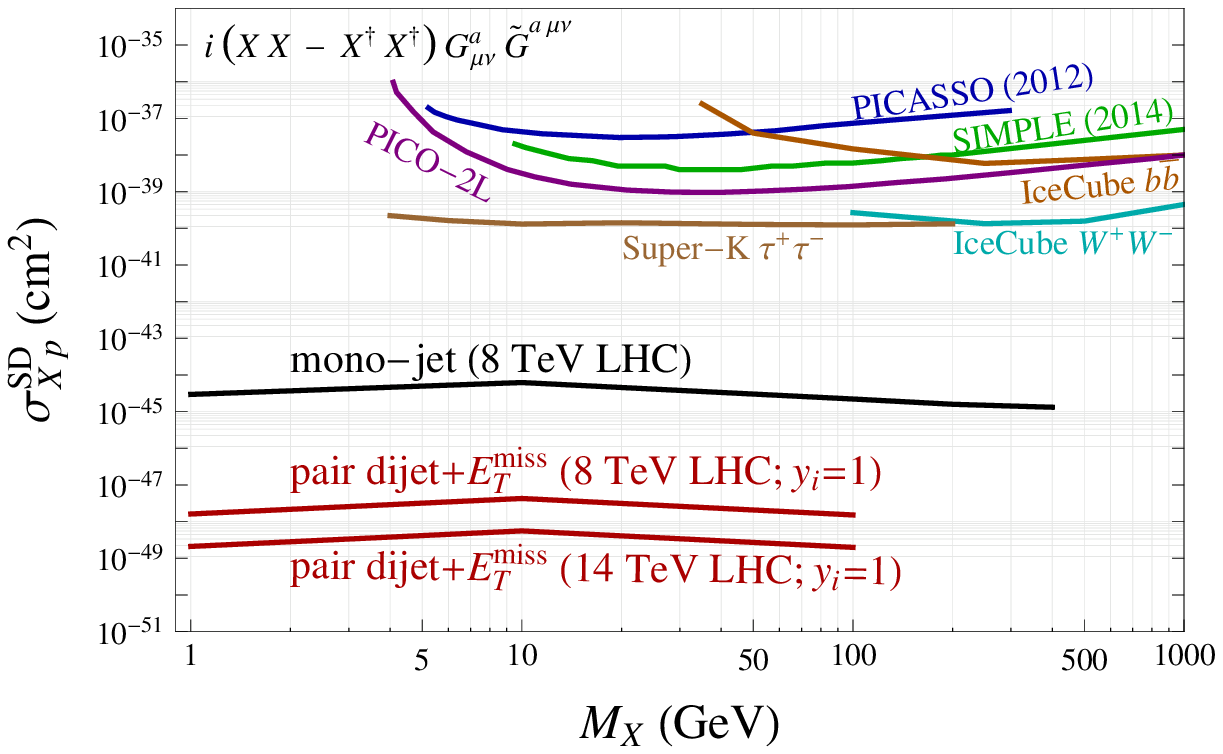} 
\caption{{\bf Left panel:} the constraints on the dark matter spin-independent scattering cross sections from the LHC mono-jet~\cite{Khachatryan:2014rra} and multi-jet+$E_T^{\rm miss}$~\cite{Chatrchyan:2014lfa} searches at the 8 TeV LHC with $\sim 20$ fb$^{-1}$ and the LUX direct detection experiment~\cite{Akerib:2013tjd}. {\bf Right panel:} the same as the left one but for spin-dependent dark matter-proton scattering cross sections from the LHC, SIMPLE~\cite{Felizardo:2014awa}, PICASSO~\cite{Archambault:2012pm}, PICO-2L~\cite{Amole:2015lsj}, IceCube~\cite{Aartsen:2012kia} and Super-K~\cite{Choi:2015ara}. The scattering cross section is suppressed by both the spin-dependent scattering and the exchanging momenta.}
\label{fig:contact-scattering}
\end{center}
\end{figure}

To compare with the constraints from direct detection experiments, we show the interpreted dark matter-nucleon scattering cross sections from the LHC mono-jet, $4j+E_T^{\rm miss}$ and paired dijet$+E_T^{\rm miss}$ searches in Fig.~\ref{fig:contact-scattering}. In the left panel, we show the spin-independent scattering cross section including the constraints from the LUX collaboration~\cite{Khachatryan:2014rra}. Compared to the direct detection limits, the constraints from the LHC are not strong for heavier dark matter masses, but are more stringent for a light dark matter mass below around 5 GeV. Compared to the limits from mono-jet searches, the $4j+E_T^{\rm miss}$ signature is definitely a more sensitive channel for constraining the spin-independent scattering cross section from the first chromo-Rayleigh interaction, $X^\dagger X G^a_{\mu\nu}G^{a\, \mu\nu}$.

 In the right panel of Fig.~\ref{fig:contact-scattering}, we show the interpreted collider constrains for the second operator, $i(XX - X^\dagger X^\dagger) G^a_{\mu\nu}\widetilde{G}^{a\,\mu\nu}$. Here, we only show the dark matter-proton scattering cross section, since it has a larger value than dark matter-neutron one for the same dark matter mass. For the interpreted limits from collider searches and because of the additional momentum suppression with $q \approx \mu_{X A} v$, we fix the dark matter velocity to be $v=10^{-3}$ and choose a typical target nucleus mass $m_A=100$~GeV in Eq.~(\ref{eq:scattering-O2}). Compared to the constraints from direct detection experiments, the collider searches obviously provide a more sensitive probe of this type of dark matter chromo-Rayleigh interactions. 

%

\section{Discussion and Conclusions}
\label{sec:conclusion}

One advantage of introducing simplified models is that the relevant model is complete from the renormalizability point of view. There is no worry about trusting theoretical descriptions of the collider limits like in the effective operator approach. On the other hand, the simplified models typically contain more model parameters and make the comparison with the results from dark matter direct or indirect detection less model-independent. For the simplified models studied here, the collider signatures highly depend on how the QCD charged particles decay. For the first operator, the color-octet scalar can decay into two gluons or two quarks plus the dark matter particle. For the second operator, the two color-triplet fermions could have many other possible decay channels beyond the ones in our paper. For instance, one could replace the operator in Eq.~(\ref{eq:psi-decay-operator}) by $\bar{\psi}_{i\, L} \sigma^{\mu\nu} u_R \widetilde{B}^{\mu\nu}$, which can lead to the decay of $\psi \rightarrow u + \gamma/Z$. The final collider signature could be $2j + 2\gamma(Z)+E_T^{\rm miss}$ with or without a pair of $j+\gamma(Z)$ resonances. To cover the majority of simplified models for dark matter chromo-Rayleigh interactions, one should search for a wide range of potential signatures. 

In conclusion, we have studied simplified models to UV-complete the chromo-Rayleigh interactions of dark matter. For the first operator, a new color-octet scalar particle could be in reach of the LHC. It may decay into just two jets or two jets plus missing transverse energy. The 8 TeV LHC can already constrain this color-octet scalar mass to be above 500-600 GeV, which can be translated into a constraint on the cutoff of the effective operator to be above 1.5-1.7 TeV. For the second operator, two QCD-charged fermions are predicted for the simplified model in this paper. A collider study of pair-produced dijet resonances plus missing energy at the 14 TeV LHC can constrain the geometric mean of the two fermion masses above around 800 GeV, which can be interpreted as a bound on the effective operator cutoff above around 1.8 TeV. 

\vspace{1cm}
{\it Note added:} We note here that during the completion of our paper, another paper~\cite{Godbole:2015gma} appeared containing some overlap with our UV-completion of the first operator, $X^\dagger X G^a_{\mu\nu}G^{a\, \mu\nu}$. 

\subsection*{Acknowledgments}
We would like to thank Ran Lu for useful discussion. This work is supported by the U. S. Department of Energy under the contract DE-FG-02-95ER40896. 

\appendix
\section{Loop Level Calculations for the Coefficient of ${\cal O}^{\rm cRayleigh}_2$}
\label{appendix}

\begin{figure}[th!]
\begin{center}
\includegraphics[width=0.48\textwidth]{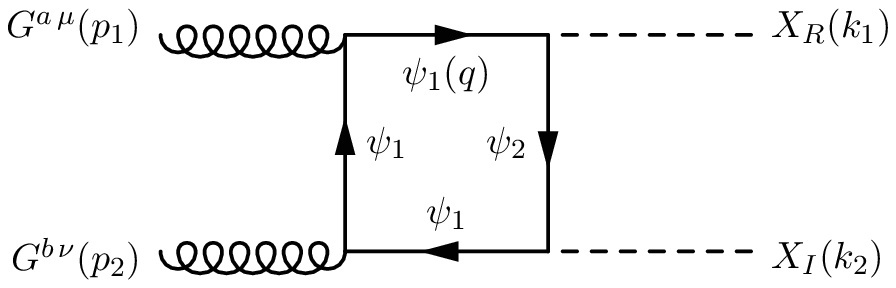} \hspace{3mm}
\includegraphics[width=0.48\textwidth]{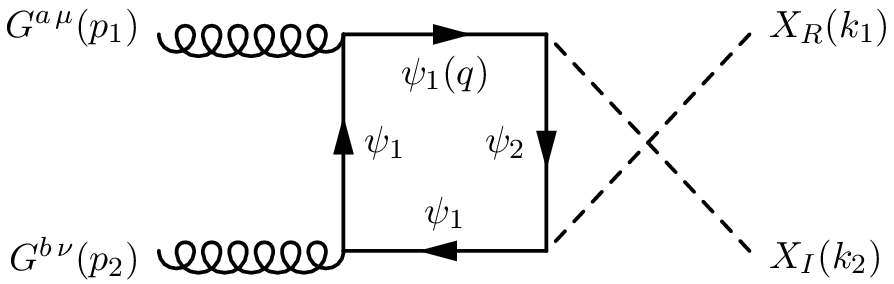} 
\caption{Loop diagrams for generating ${\cal O}^{\rm cRayleigh}_2$ via a colored particle in loop..}
\label{fig:box-diagram-24}
\end{center}
\end{figure}

\begin{figure}[th!]
\begin{center}
\includegraphics[width=0.48\textwidth]{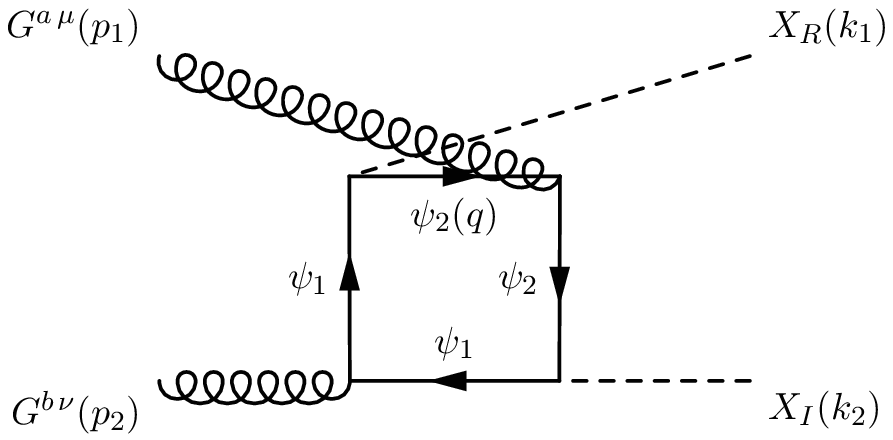} \hspace{3mm}
\includegraphics[width=0.48\textwidth]{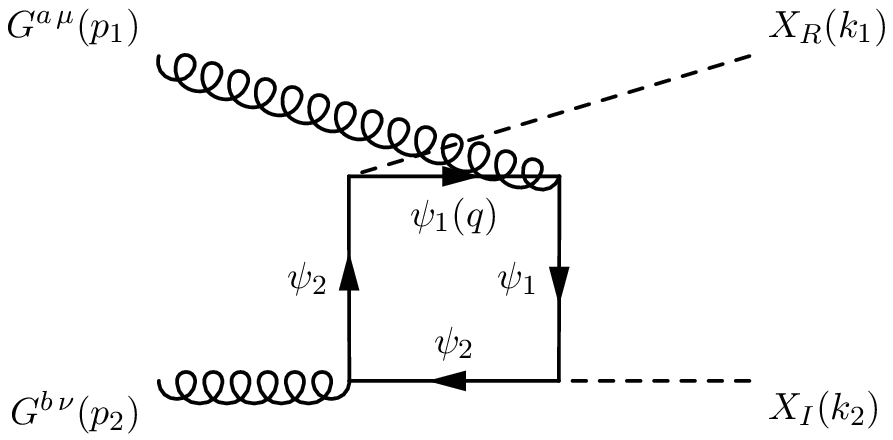} 
\caption{Loop diagrams for generating ${\cal O}^{\rm cRayleigh}_2$ via a colored particle in loop..}
\label{fig:box-diagram-56}
\end{center}
\end{figure}

Starting from the Lagrangian in Eq.~(\ref{eq:XI-lagrangian}), we match the coefficient for ${\cal O}^{\rm cRayleigh}_2$ at one-loop level. Other than the two Feynman diagrams in Fig.~\ref{fig:box-diagram-example}, we also have another four diagrams in Fig.~\ref{fig:box-diagram-24} and Fig.~\ref{fig:box-diagram-56}.

Summing the two diagrams in Fig.~\ref{fig:box-diagram-example}, we have the matrix element as
\beqa
  i \mathcal{M}_1 &= 4i g_s^2 y_1 y_2 \text{Tr} [ t^a t^b ] \varepsilon_\mu(p_1) \varepsilon_\nu(p_2) 
 \times \int \frac{d^4 q}{(2 \pi)^4} \frac{\text{Tr} [ (\slashed{q} + m_2) \gamma^\mu (\slashed{q} - \slashed{p}_1 + m_2) \gamma^\nu (\slashed{q} - \slashed{k}_1 - \slashed{k}_2 + m_2) \gamma^5 (\slashed{q} - \slashed{k}_1 + m_1) ]}{[q^2 - m_2^2] [(q - p_1)^2 - m_2^2] [(q - k_1 - k_2)^2 - m_2^2] [(q - k_1)^2 - m_1^2]} \, .
\eeqa
In the heavy fermion limit, we have a simplified version of the matrix element as
\beqa
i \mathcal{M}_1 =  i\, \frac{g_s^2 y_1 y_2}{4\pi^2} \delta^{ab} \varepsilon_\mu(p_1) \varepsilon_\nu(p_2) \epsilon^{\mu \nu \rho \sigma} p_{1\rho} p_{2\sigma} \left[ \frac{ - m_1^5 + 2 m_1^3 m_2^2 \log \left ( m_1^2 / m_2^2 \right ) + m_1 m_2^4 }{m_2 (m_1^2 - m_2^2)^3} \right] \,. 
\eeqa
The two diagrams in Fig.~\ref{fig:box-diagram-24} have a similar answer as
$i \mathcal{M}_2 = i \mathcal{M}_1 (m_1 \leftrightarrow m_2)$. The two digrams in Fig.~\ref{fig:box-diagram-56} has a summed matrix as
\beqa
  i \mathcal{M}_3 =i \frac{g_s^2 y_1 y_2}{2\pi^2} \delta^{ab} \varepsilon_\mu(p_1) \varepsilon_\nu(p_2) \epsilon^{\mu \nu \rho \sigma} p_{1\rho} p_{2\sigma} \left\{\frac{ m_1 m_2 \left [ 2 (m_1^2 - m_2^2 ) - ( m_1^2 + m_2^2 ) \log \left ( m_1^2 / m_2^2 \right ) \right ]}{ (m_1^2 - m_2^2)^3} \right\} \,.
\eeqa
Add all diagrams together, we have the summed matrix element
\beqa
i \mathcal{M}_1 + i \mathcal{M}_2 + i \mathcal{M}_3 = \frac{g_s^2 y_1 y_2}{4\pi^2\,m_1\,m_2} \delta^{ab} \varepsilon_\mu(p_1) \varepsilon_\nu(p_2) \epsilon^{\mu \nu \rho \sigma} p_{1\rho} p_{2\sigma} \,.
\eeqa
To match to the coefficient of ${\cal O}^{\rm cRayleigh}_2$, one has
\beqa
\frac{\alpha_s\, y_1 y_2}{\pi\,m_1\,m_2}  = \frac{2\,\alpha_s}{\pi\,\Lambda^2_2} \,.
\eeqa

\bibliography{DMCRayleigh}
\bibliographystyle{JHEP}
 \end{document}